\documentclass[lettersize,journal]{IEEEtran}
\usepackage{amsmath,amsfonts}
\usepackage{algorithmic}
\usepackage{algorithm}
\usepackage{array}
\usepackage[caption=false,font={rm,scriptsize}]{subfig}
\usepackage{textcomp}
\usepackage{stfloats}
\usepackage{xurl}
\usepackage[
    breaklinks=true,
    hidelinks
]{hyperref}
\usepackage{verbatim}
\usepackage{graphicx}
\usepackage{cite}
\usepackage{siunitx}

\usepackage{booktabs}  
\usepackage{multirow}  
\usepackage{makecell}  

\usepackage{xcolor}
\newcommand{\rev}[1]{{#1}}

\begin{document}


\title{Towards Ultra-High Reliability in Wi-Fi 8: IEEE 802.11bn Core Mechanisms, mmWave Integration, and Performance Verification}

\author{Xiaoqian Liu,~\IEEEmembership{Graduate Student Member,~IEEE,}
        Ming Gan,~\IEEEmembership{Member,~IEEE,}\\
        Weijie Dai,~\IEEEmembership{Graduate Student Member,~IEEE,}
        Yuhan Dong,~\IEEEmembership{Senior Member,~IEEE,}\\
        Calvin Chun-Kit Chan,~\IEEEmembership{Senior Member,~IEEE, Fellow,~Optica,}
        and~Jian Song,~\IEEEmembership{Fellow,~IEEE}

\thanks{Xiaoqian Liu and Calvin Chun-Kit Chan are with the Department of Information Engineering, The Chinese University of Hong Kong, Hong Kong (e-mail: lx025@ie.cuhk.edu.hk; ckchan@ie.cuhk.edu.hk).}
\thanks{Ming Gan is with the Wireless Technology Lab, Huawei Technologies Company Ltd., Shenzhen 518129, China (e-mail: ming.gan@huawei.com).}
\thanks{Weijie Dai, Yuhan Dong and Jian Song are with the Tsinghua Shenzhen International Graduate School, Shenzhen 518055, China (e-mail: \urlstyle{same}\href{mailto:daiwj1999@outlook.com}{\nolinkurl{daiwj1999@outlook.com}}; \href{mailto:dongyuhan@sz.tsinghua.edu.cn}{\nolinkurl{dongyuhan@sz.tsinghua.edu.cn}}; \href{mailto:jsong@tsinghua.edu.cn}{\nolinkurl{jsong@tsinghua.edu.cn}}).}

\thanks{\textit{Corresponding authors: Yuhan Dong and Ming Gan.}}

}



\maketitle

\begin{abstract}
As the demand for wireless connectivity expands from high-speed data transmission to high-reliability applications, such as the Industrial Internet of Things and immersive communications, traditional Wi-Fi technologies optimized primarily for peak throughput face new challenges in reliability and latency.
Consequently, Wi-Fi 8 aims to achieve ultra-high reliability (UHR), improve communication performance in complex environments, and drive the transition from high-speed connectivity to highly reliable intelligent connectivity.
This article provides a comprehensive review of the core mechanisms of Wi-Fi 8 and conducts system-level performance verification.
We focus on the key enhancement mechanisms at the physical (PHY) and medium access control (MAC) layers in IEEE 802.11bn, elaborating on their theoretical principles and key application scenarios.
Additionally, this paper explores the \rev{potential} role of integrated millimeter-wave (IMMW) technology \rev{as a complementary solution for spectrum expansion in the Wi-Fi 8 era}, analyzing its basic architecture and implementation. 
Finally, system-level simulations are performed to verify the effectiveness of the key technologies in IEEE 802.11bn in achieving their performance targets, while further validating the robust performance of the IMMW scheme under practical hardware impairments.

\end{abstract}


\section{Introduction}
\IEEEPARstart{O}{ver} the past quarter-century, Wi-Fi has seamlessly transitioned from a technological convenience to the invisible backbone of modern society \cite{wifi25, wifi8powersave}.
Today, connecting to a local wireless network has become as instinctive to the modern citizen as switching on the lights upon entering a room.
Serving as the primary conduit for global connectivity, this technology currently shoulders the vast majority of wireless data traffic across tens of billions of active devices \cite{willwifi8be}. It profoundly reshapes daily routines by sustaining uninterrupted cloud-based enterprise collaborations, animating smart domestic ecosystems, and democratizing high-definition entertainment.
As its footprint continually expands, Wi-Fi is gradually outgrowing its historical role as a mere data pipe, laying the foundation for a broader spectrum of intelligent, context-aware services \cite{comstsensing}.

Although Wi-Fi 7 can achieve peak throughput rates of up to 30 Gbps \cite{Liuwifi7, wifi72021back, wifi72022, IEEE80211be, 19wifi7}, increasing peak data rates alone is insufficient to meet the demands of future networks.
Emerging applications, such as the Industrial Internet of Things (IIoT), \rev{Physical artificial intelligence (AI),} collaborative mobile robotics, and immersive communications, impose stringent requirements on deterministic latency and high reliability \cite{2022reviewcommmag}.
These strict demands exceed the capabilities of existing wireless local area network (WLAN) standards, which still struggle with unpredictable delays and insufficient transmission reliability.
Furthermore, the escalating spectrum congestion within traditional sub-7 GHz bands presents a significant challenge, prompting the exploration of the millimeter-wave (mmWave) spectrum as a critical alternative \cite{IEEE80211bqPAR, LiuIMMW}.
To address these issues, the IEEE 802.11 working group has initiated two key research efforts. 
The IEEE 802.11bn amendment, known as Ultra High Reliability (UHR), aims to enhance reliability, latency performance, and network resilience through modifications to the physical (PHY) and medium access control (MAC) layers \cite{IEEE80211bnPAR}.
Concurrently, the Integrated Millimeter Wave (IMMW) task group (IEEE 802.11bq) focuses on \rev{developing mmWave links as a complementary extension to} the Wi-Fi 8 framework, utilizing existing low-frequency designs to enable seamless cross-band coordination \cite{LiuIMMW, IEEE80211bqPAR}.
Together, these standard evolutions mark a paradigm shift for Wi-Fi, transitioning from the historical focus on peak speeds to a new era centered on robust and deterministic reliability. 

In this article, we provide a comprehensive review of the core mechanisms of Wi-Fi 8 and evaluate their performance through system-level simulations.
We first review the evolutionary trajectory of Wi-Fi standards and highlight the primary innovations across previous generations.
Subsequently, we examine the key enhancements at the PHY and MAC layers introduced by IEEE 802.11bn.
Furthermore, we discuss the cross-band coordination architecture based on integrated mmWave technology.
System-level simulations are then presented to verify the effectiveness of these technologies in achieving the key performance objectives of Wi-Fi 8.
By combining theoretical analysis with empirical validation, this article aims to help readers better understand the fundamental technologies of Wi-Fi 8 and provide a reliable reference for future research in this field.

\begin{figure*}[!t]
    \centering
    \includegraphics[width=\textwidth]{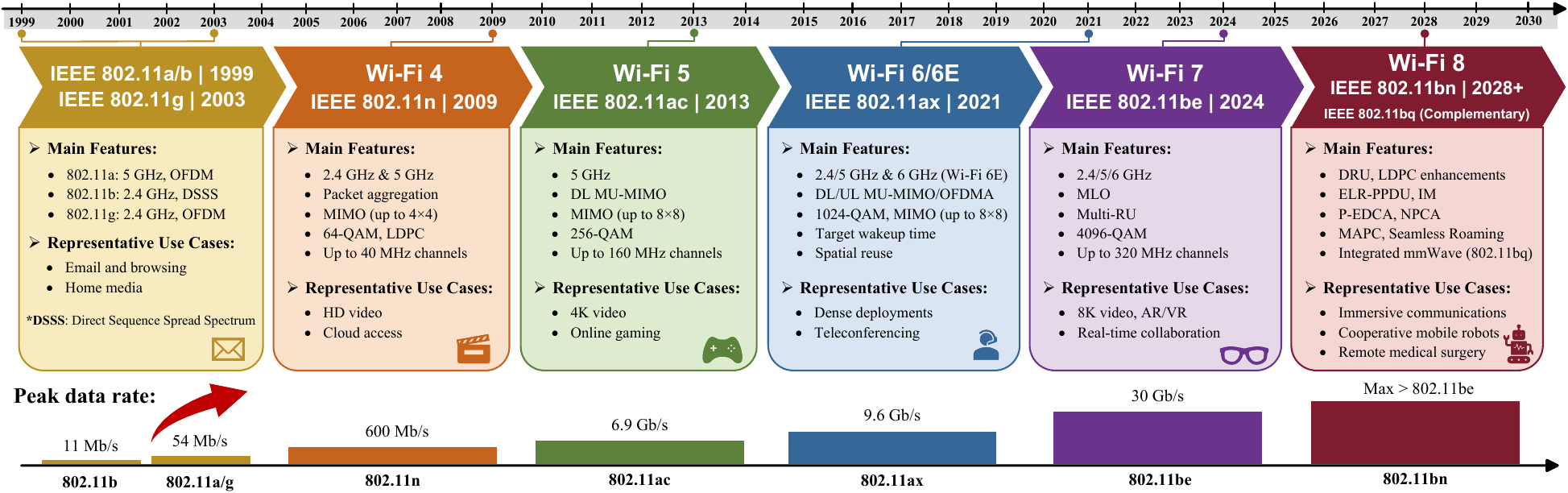}
    \caption{Comprehensive timeline of Wi-Fi standards evolution, illustrating the technological progression from legacy IEEE 802.11a to the upcoming Wi-Fi 8 era. The figure details standard designations, key features, peak data rates, and representative use cases for each generation.}
    \label{fig:wifi_timeline}
\end{figure*}

\subsection{The Quest for Capacity: Evolution from Legacy Standards to Wi-Fi 6}
\rev{The historical trajectory of Wi-Fi has been primarily driven by the continuous demand for higher data rates.}
Fig.~\ref{fig:wifi_timeline} illustrates the evolution timeline of Wi-Fi standards and highlights their major features.
Following its initial release in 1997 with a modest 2 Mbps peak rate, early amendments such as IEEE 802.11b, 802.11a, and 802.11g gradually increased data rates to 54 Mbps \cite{wifi25}. This improvement was largely achieved through the adoption of orthogonal frequency-division multiplexing (OFDM), which laid the foundation for basic wireless internet access and email connectivity.

A significant architectural advancement occurred in 2009 with the introduction of Wi-Fi 4 (IEEE 802.11n) \cite{WiFiAllianceN}. By implementing multiple-input multiple-output (MIMO) technology, channel bonding, \rev{and low-density parity-check (LDPC) coding}, Wi-Fi 4 reached peak rates of 600 Mbps, facilitating bandwidth-intensive applications such as high-definition video streaming \cite{80211n}. Subsequently, Wi-Fi 5 (IEEE 802.11ac) was certified in 2013, operating primarily in the 5 GHz band \cite{WiFiAllianceAC}. It introduced wider channels up to 160 MHz, higher-order modulation (256-QAM), and downlink multi-user MIMO (MU-MIMO) \cite{80211ac}. These features enabled multi-gigabit data rates, cementing Wi-Fi as the primary medium for enterprise networking and mobile data offloading.

As device density escalated in public and residential spaces, the focus of standard development transitioned from peak data rates to spectral efficiency. Released in 2021, Wi-Fi 6 (IEEE 802.11ax) introduced orthogonal frequency-division multiple access (OFDMA) and spatial reuse techniques to optimize channel allocation and reduce network contention \cite{WiFiAlliance6}. Supporting theoretical rates up to 9.6 Gbps, Wi-Fi 6 significantly improved average per-user throughput in congested \rev{and dense} environments \cite{80211ax}. Shortly after, Wi-Fi 6E extended these capabilities into the unlicensed 6 GHz band, providing additional spectrum free from legacy device interference and accommodating the growing volume of modern wireless traffic \cite{LukaszewskiWiFi6E}.
\subsection{The Era of Extreme Throughput: Reaching the Limits with Wi-Fi 7}
Certified in 2024, the IEEE 802.11be amendment, commercially designated as Wi-Fi 7, was developed under the explicit objective of delivering Extremely High Throughput (EHT) \cite{Liuwifi7}. To fulfill this requirement, the standard introduced substantial enhancements across both the PHY and MAC layers. 
At the physical layer, Wi-Fi 7 doubled the maximum channel bandwidth to 320 MHz within the 6 GHz spectrum, incorporated 4096-QAM to increase data density per transmission symbol, and introduced support for multiple resource units (MRUs) in OFDMA. Supported by MAC-layer scheduling, this flexible resource unit (RU) assignment enhances spectral efficiency by providing more flexible resource allocation to individual stations \cite{wifi72022}.

At the MAC layer, the most significant innovation is multi-link operation (MLO), which fundamentally shifts the legacy single-link paradigm. MLO enables a device to aggregate traffic, transmit concurrently, or dynamically switch across the 2.4 GHz, 5 GHz, and 6 GHz bands \cite{WIFI7MLO}. This architectural capability significantly boosts aggregate throughput, reduces communication latency by increasing channel access opportunities, and improves overall connection reliability through transmission redundancy. By synergizing these features, Wi-Fi 7 achieved theoretical peak data rates exceeding 30 Gbps, successfully accommodating bandwidth-intensive applications such as 8K video streaming and pushing single-access-point performance to unprecedented levels.

\subsection{Towards Low Latency and High Reliability: The Vision for Wi-Fi 8}
Many users have experienced the frustration of failing to connect to a Wi-Fi network precisely when it is needed most, or encountering delays and stuttering during video conferences. Unreliability is a critical vulnerability for affordable and widely deployed technologies operating in interference-prone open spectrum bands, and Wi-Fi is no exception. While human users may patiently tolerate video buffering or repeat sentences during a stalled voice call, automated systems and machines absolutely cannot accept insufficient Wi-Fi reliability. In remote medical surgeries, an unreliable connection directly threatens patient safety. Similarly, in immersive holographic communications, excessive delay in merely 0.01 percent of transmitted packets can trigger user discomfort and motion sickness \cite{willwifi8be}. Furthermore, future indoor connectivity scenarios will rely on Wi-Fi for mission-critical tasks such as digital twinning and collaborative mobile robotics, all of which mandate reliability levels of 99.9 percent or higher \cite{ojcomreview}. Driven by the stringent latency and reliability demands of these emerging applications, the next generation of Wi-Fi will prioritize Ultra High Reliability (UHR) \cite{ComputerNetworkswifi7to8}. The IEEE 802.11bn amendment is explicitly tasked with addressing these challenges and will form the technological foundation for Wi-Fi 8.

The ambition to improve network latency initially emerged within the IEEE 802.11be (Wi-Fi 7) Project Authorization Request (PAR) \cite{IEEE80211bePAR}. Although the Wi-Fi 7 PAR primarily targeted a maximum throughput of 30 Gbps, it also introduced the goal of improving worst-case latency and jitter. However, these latency objectives were not quantified, serving only as a preliminary step toward reliable connectivity. 
The IEEE 802.11bn PAR addresses this limitation by establishing low latency and high reliability as strictly quantifiable targets \cite{IEEE80211bnPAR}. Specifically, the 802.11bn amendment mandates at least one mode of operation that delivers a 25 percent throughput increase at specific signal-to-interference-plus-noise ratio (SINR) levels, which directly enhances rate-vs-range performance. Furthermore, it requires a 25 percent reduction in both the 95th percentile of the latency distribution and the MAC protocol data unit (MPDU) loss rate, particularly during transitions between overlapping networks. This evolution from qualitative latency ambitions to precise performance guarantees signifies a fundamental shift in network design, ensuring that Wi-Fi 8 can fulfill its ultra-high reliability mandate.

To achieve these stringent UHR objectives, Wi-Fi 8 introduces a series of critical technological innovations.
Among them, multi-access point coordination (MAPC) will be one of the key areas of development \cite{Wi-Fi8Unveiled, Karamyshevreview, wilhelmi2026tutorialieee80211bnmultiap}.
MAPC facilitates coordination among neighboring access points (APs) to decrease collisions and improve medium usage efficiency \cite{comstreview}. 
Moving beyond traditional mesh network architectures, it supports coordinated scheduling across multiple APs to proactively manage interference and channel access. 
This approach shifts Wi-Fi away from its historical reliance on purely nondeterministic contention mechanisms, laying the groundwork for reliable performance in highly dense deployments.
Also within the MAC layer, seamless roaming technology is introduced to enable uninterrupted connectivity and efficient operations during roaming \cite{seamlessroaming}. 
By facilitating fast handovers across APs, it directly supports mobility-sensitive applications such as augmented and virtual reality (AR/VR) and real-time industrial control. 
At the PHY layer \cite{ojcomreview}, the extended long range (ELR) technology provides essential range improvement and enhances transmission reliability, particularly in environments with high interference. 
Furthermore, distributed resource units (DRUs) address power spectral density (PSD) limitations imposed by specific regulatory regimes. 
By increasing power distribution flexibility, DRUs allow for improved energy concentration and transmission robustness. 
\rev{Additionally, interference mitigation (IM) provides an effective solution to suppress interference, significantly increasing goodput at specific SINR levels.}
Beyond these core mechanisms, the standard encompasses several other enhancements, including prioritized enhanced distributed channel access (P-EDCA), non-primary channel access (NPCA) \cite{npca}, and dynamic subband operation (DSO).

Beyond the optimizations in the IEEE 802.11bn amendment, the IEEE 802.11 working group is concurrently exploring the mmWave spectrum through the 802.11bq IMMW project \cite{LiuIMMW, IEEE80211bqPAR}.
As traditional sub-7 GHz bands face increasing congestion and regulatory bandwidth limits, the 60 GHz band provides abundant contiguous spectrum to satisfy escalating capacity demands.
To achieve this, the IMMW approach intends to reuse existing sub-7 GHz physical layer designs to minimize hardware complexity and production costs.
Furthermore, by extending the MLO framework, devices can maintain robust control connections over sub-7 GHz frequencies while offloading heavy data payloads to high-throughput mmWave links \cite{yuan2026throughputoptimizationmultiapieee}.
This cross-band synergy effectively complements the reliability objectives \rev{in the Wi-Fi 8 era.}

\begin{figure}[!t]
    \centering
    \includegraphics[width=\columnwidth]{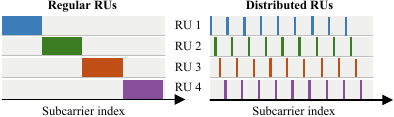}
    \caption{Comparison of subcarrier allocation between RRU and DRU.}
    \label{fig:dru_allocation}
\end{figure}

\section{Fortifying the Physical Foundation: Link Robustness and Interference Mitigation}
\subsection{Distributed Resource Unit (DRU)}
In the evolution of Wi-Fi up to the IEEE 802.11be standard, OFDMA exclusively utilized continuous subcarriers to form RUs \cite{wifi25}.
While efficient for general traffic multiplexing, this contiguous allocation encounters a fundamental constraint in shared spectrum scenarios, particularly within the 6 GHz band governed by low-power indoor (LPI) regulations \cite{DRULowP}.
Under LPI rules, device transmissions are strictly limited by a PSD threshold rather than an absolute total power limit.
When a station (STA) transmits using a small, continuous RU, such as a 26-tone RU spanning approximately 2 MHz, its total allowable transmit power is severely restricted by this narrow bandwidth.
This physical bottleneck degrades the uplink link budget and significantly compromises the performance for devices operating at the cell edge.

To resolve this limitation, the IEEE 802.11bn amendment introduces the DRU \cite{DRUproposal}.
The core mechanism of the DRU relies on frequency-domain scattering \cite{DRUTUNEPLAN1}.
As shown in Fig.~\ref{fig:dru_allocation}, instead of grouping subcarriers adjacently, a DRU strategically scatters nonadjacent tones evenly across a wider operational bandwidth, defined as the distributed bandwidth (DBW).
By reducing the density of subcarriers within any single 1 MHz segment, the STA is permitted to concentrate more energy into each individual tone.
Consequently, the device can effectively increase its total transmit power while remaining strictly compliant with regulatory PSD constraints, yielding substantial uplink power gains.
By uniformly dispersing tones across the operational bandwidth, DRUs sustain a comparable user capacity per channel as conventional OFDMA without compromising spectral efficiency, thereby ensuring that diverse traffic demands are managed effectively \cite{DRUTUNEPLAN2, DRUTUNEPLAN3}.

Beyond overcoming power limitations, the uniform distribution of tones inherently provides significant frequency diversity. This characteristic makes DRU-based transmissions notably more robust in environments with high channel fading, as they become highly resilient to frequency-selective fading in complex multipath conditions.
Furthermore, to minimize implementation complexity, the standard preserves the hierarchical tone mapping structure and pilot tone placement used in regular RUs (RRUs), which directly ensures compatibility and ease of implementation.
Despite these compelling advantages, scattering tones across a wide spectrum introduces specific engineering trade-offs. For example, the disruption of continuous subcarrier alignment elevates the peak-to-average power ratio (PAPR) of the transmitted signal \cite{DRUPAPR}.
To address this, the standardization task group is actively optimizing the ultra high reliability long training field (UHR-LTF) sequences using precise phase rotations to suppress time-domain signal peaks \cite{DRUltf}.

\begin{figure}[!t]
    \centering
    \subfloat[Four-fold frequency domain duplication mechanism]{
        \includegraphics[width=\columnwidth]{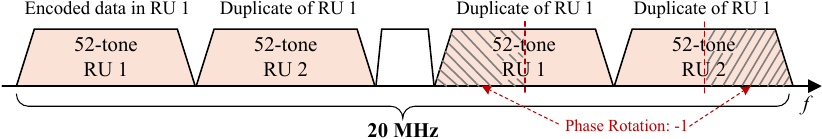}
        \label{fig:elr_a}
    }
    
    \vspace{-5pt}
    
    \subfloat[UHR ELR PPDU]{
        \includegraphics[width=\columnwidth]{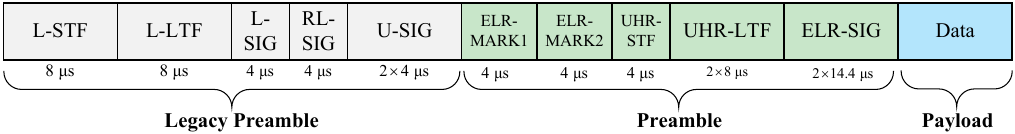}
        \label{fig:elr_b}
    }
    
    \caption{Overview of the ELR feature in IEEE 802.11bn.}
    \label{fig:elr_overview}
\end{figure}

\subsection{Enhanced Long Range (ELR)}
To extend coverage and ensure connection stability at the network edge, the IEEE 802.11bn amendment introduces the enhanced long range (ELR) physical layer protocol data unit (PPDU) \cite{elrconsider, elrOPENTOPIC}.
The primary motivation behind this design is to resolve the inherent link budget asymmetry between the downlink and uplink channels.
This asymmetry arises because APs generally transmit at significantly higher power levels than STAs, meaning that uplink signals often fail to reach the receiver in edge scenarios \cite{elrpowerNON}.

According to the protocol specifications, the ELR format supports bidirectional communications in the 2.4 GHz band, but is strictly restricted to uplink transmissions in the 5 GHz and 6 GHz bands. The ELR PPDU prioritizes maximum robustness over spectral efficiency. It is confined to a 20 MHz channel bandwidth and a single spatial stream.
Furthermore, it only utilizes the most resilient modulation and coding schemes (MCSs), specifically ELR MCS 0 and 1, which correspond to binary phase shift keying (BPSK) and quadrature phase shift keying (QPSK), \rev{both with a 1/2 coding rate.}
As shown in Fig.~\ref{fig:elr_overview}(a), the core mechanism of this transmission mode is a four-fold frequency domain duplication \rev{with segment rotation}.
A single encoded payload is identically replicated and mapped across four separate 52-tone RUs within the 20 MHz channel \cite{elrtx}. 
To suppress the PAPR increase inherently caused by waveform duplication, a phase rotation of -1 is applied directly to the lower half of the third 52-tone RU and the upper half of the fourth 52-tone RU \cite{elrphase}.
By combining deep redundancy and robust modulation, ELR significantly enhances uplink coverage and transmission reliability, making it particularly advantageous for deployment scenarios such as outdoor campuses and industrial facilities.

As illustrated in Fig.~\ref{fig:elr_overview}(b), the ELR PPDU frame structure is designed to maintain backward compatibility while accommodating these new features \cite{elrppdu}.
The preamble begins with the legacy fields identical to those in the EHT PPDU, up to the universal signal field (U-SIG).
Following the U-SIG, the frame introduces two ELR-MARK symbols primarily utilized for ELR mode classification.
This is followed by the UHR short and long training fields (UHR-STF and UHR-LTF) for synchronization and channel estimation.
Finally, an ELR-SIG field is inserted before the data payload to convey crucial demodulation parameters, such as the transmission direction and the selected MCS.

\begin{figure}[!t]
    \centering
    \subfloat[P-EDCA]{
        \includegraphics[width=\columnwidth]{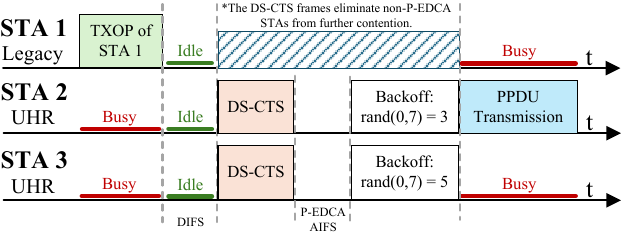}
        \label{fig:op_pedca}
    }
    
    \vspace{-5pt}
    
    \subfloat[NPCA]{
        \includegraphics[width=\columnwidth]{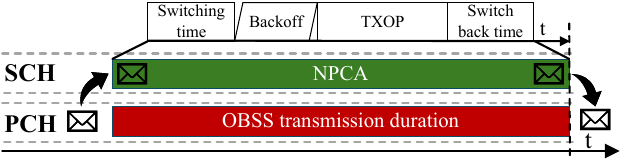}
        \label{fig:op_npca}
    }
    
    \caption{Operational procedures of the P-EDCA and NPCA mechanisms. DIFS denotes the distributed coordination function interframe space.}
    \label{fig:pedca_npca_operations}
\end{figure}

\subsection{Interference Mitigation (IM)}
Traditional rate adaptation mechanisms select an appropriate MCS based on current channel conditions.
However, this approach is generally unable to effectively handle sudden and rapidly time-varying interference.
To address this limitation, the IEEE 802.11bn amendment introduces interference mitigation (IM) to help multi-antenna STAs detect and suppress interference in real time, thereby further improving transmission reliability.

The core of the IM technique involves embedding additional pilot tones within the data portion of the PPDU \cite{IM1, IMPOLIT}.
These specific pilots enable the receiver to calculate the statistical characteristics of channel noise and burst interference.
When the number of physical receive antennas exceeds the number of transmitted spatial streams, the receiver possesses redundant spatial degrees of freedom.
In this scenario, the receiver can utilize the interference features extracted from the IM pilots to configure receiver-side beamforming \cite{IM2}. 
Consequently, the application of IM allows the system to maintain reliable connectivity even under strong interference environments.

Nevertheless, the application of IM introduces specific trade-offs \cite{IM1, IMconsider}. 
Inserting denser pilot tones improves interference suppression but inevitably reduces the number of subcarriers available for carrying actual payload data.
Additionally, a trade-off exists between the flexibility of pilot pattern configurations and implementation complexity. To balance adaptability and signaling overhead, current standardization efforts tend to favor a limited set of predefined, flexible pilot configurations.

\section{Enhancing Channel Access: Prioritization, Flexibility, and Latency Reduction}
\subsection{Prioritized Enhanced Distributed Channel Access (P-EDCA)}
The original IEEE 802.11 distributed coordination function (DCF) relied on basic carrier sense multiple access with collision avoidance (CSMA/CA).
To support quality of service, the IEEE 802.11e amendment introduced enhanced distributed channel access (EDCA), which classifies traffic into four access categories and assigns the highest priority to voice traffic (AC\_VO) \cite{80211eedca}. 
While EDCA effectively differentiates traffic when high-priority packets are sparse, its probabilistic nature struggles under heavy network loads.
In dense deployments, severe contention among multiple AC\_VO flows leads to frequent collisions and unacceptable delay spikes. This makes a more deterministic approach essential for delay-sensitive applications.

To resolve these latency bottlenecks, the IEEE 802.11bn amendment defines the prioritized EDCA (P-EDCA) mechanism \cite{pedcaconsider}.
This feature is specifically designed to truncate the tail of the access delay distribution for highly urgent traffic \cite{pedcadiscussion}.
Rather than applying to all transmissions, P-EDCA is conditionally activated. The primary trigger occurs when the retransmission counter for an AC\_VO frame reaches a predefined threshold, indicating multiple consecutive delivery failures.
Upon meeting this criterion, the STA is granted exclusive P-EDCA privileges, allowing it to temporarily bypass conventional contention protocols.

Fig.~\ref{fig:pedca_npca_operations}(a) illustrates the operational procedure of P-EDCA.
The operational sequence begins when the privileged STA detects a brief channel idle period.
It immediately broadcasts a short control frame known as the defer signal clear-to-send (DS-CTS) \cite{pedcadts}.
The transmission of this signal establishes an absolute priority advantage.
When non-P-EDCA STAs detect the DS-CTS, their ongoing backoff processes are instantly interrupted and suspended.
Consequently, the channel is effectively cleared, isolating the contention process exclusively to the subset of STAs that successfully transmitted a DS-CTS. These prioritized devices then engage in a dedicated short contention period.
They wait for a specialized arbitration interframe space (P-EDCA AIFS), randomly initialize a smaller backoff counter, and count down to initiate their frame exchanges \cite{pedcaframe, pedcatxop}.
Once a prioritized STA successfully acquires the channel, it must execute a request-to-send and clear-to-send (RTS/CTS) handshake.
This step protects the secured transmission opportunity (TXOP), ensuring uninterrupted transmission during this period.

\subsection{Non-Primary Channel Access (NPCA)}
Since the introduction of channel bonding \cite{CHANNELBOUNDING}, the available channel bandwidth in Wi-Fi networks has progressively expanded.
However, traditional channel access mechanisms remain strictly constrained by the mandatory designation of a 20 MHz primary channel (PCH).
Under standard operations, a device can only extend its transmission into available secondary channels (SCHs) if the PCH is sensed idle.
This requirement exposes a critical limitation in dense deployments.
If the PCH is occupied by an overlapping basic service set (OBSS), the device is completely prevented from initiating any transmission, even if a broad expanse of SCH spectrum remains perfectly idle.
This phenomenon leads to severe spectrum waste and increases channel access delays.
To address this issue, the IEEE 802.11bn amendment introduces non-primary channel access (NPCA) \cite{npcaDETAILS1, npcaDETAILS2}.
The core objective of this mechanism is to permit data transmission over idle SCHs when the traditional PCH is blocked \cite{npcaconsider}.
To facilitate this, each BSS supporting this feature pre-defines an auxiliary channel within its operating bandwidth, designated as the NPCA PCH.
Fig.~\ref{fig:pedca_npca_operations}(b) illustrates the channel switching procedure of NPCA.

For NPCA to function effectively, the AP must broadcast a set of critical parameters to its STAs \cite{npcaper}. 
These parameters specify the exact location of the NPCA PCH, the minimum duration threshold required to trigger a channel switch, and the specific time delays required to switch to the \rev{NPCA PCH} and subsequently switch back.
\rev{Additionally, the AP coordinates medium synchronization by announcing parameters such as the TXOP limit to bound the transmission time on the non-primary channel. This ensures that NPCA transmissions conclude before the PCH becomes available again, enabling the BSS to resume operation on its default PCH.}
The channel switch operates as a passive response mechanism. 
If an STA detects an OBSS PPDU transmission or an RTS/CTS control frame exchange on the PCH, it extracts the anticipated duration of this communication. If this duration exceeds the minimum duration threshold previously announced by the AP, an NPCA transition is triggered.
Upon switching to the new channel, the STA must execute the standard backoff procedure on the new channel. Once the STA successfully wins the contention on the NPCA PCH, it initiates the communication by transmitting a critical control frame known as the initial control frame (ICF). The primary function of the ICF is to reliably establish channel access and perform data initialization between the STA and the AP over this non-primary channel \cite{npcaDETAILS3}.

By allowing STAs to dynamically migrate to non-primary channels when the main channel is busy, NPCA mitigates the rigid constraints of traditional channel bonding. This capability provides an innovative and highly effective solution for maximizing spectrum utilization in high-density wireless networks \cite{npcaadvantages, npcagood}.

\subsection{In-Device Coexistence and Unavailability Reporting}
In traditional Wi-Fi networks, the AP remains unaware of internal resource contention within a connected STA. To minimize the physical size of mobile devices, Wi-Fi STAs frequently share radio frequency components with other wireless technologies, such as Bluetooth. This hardware integration introduces severe in-device coexistence (IDC) interference \cite{IDCintro}.
To resolve this limitation, the IEEE 802.11bn amendment introduces periodic unavailability operation (PUO) and dynamic unavailability operation (DUO). The core objective of these mechanisms is to empower the STA with the ability to explicitly inform the AP about its exact unavailable time intervals \cite{IDCdo}.

PUO is specifically designed to manage predictable and periodic internal interference, such as continuous Bluetooth audio streaming \cite{IDCpuo}. This mechanism is \rev{achieved} through a modified version of the target wake time (TWT) protocol. Upon receiving a PUO report, the AP adjusts its scheduling to proactively avoid transmitting frames to that specific STA during the reported periodic blind spots.
Conversely, DUO is introduced to handle unpredictable or urgent unavailability events \cite{IDCDuo}. Under the DUO mode, a STA can dynamically indicate its unavailability start time and duration either at the beginning of or during a TXOP using modified frames. This prevents unnecessary packet loss caused by sudden internal hardware contention.

\begin{figure}[!t]
    \centering
    \subfloat[Co-SR]{
        \includegraphics[width=0.8\columnwidth]{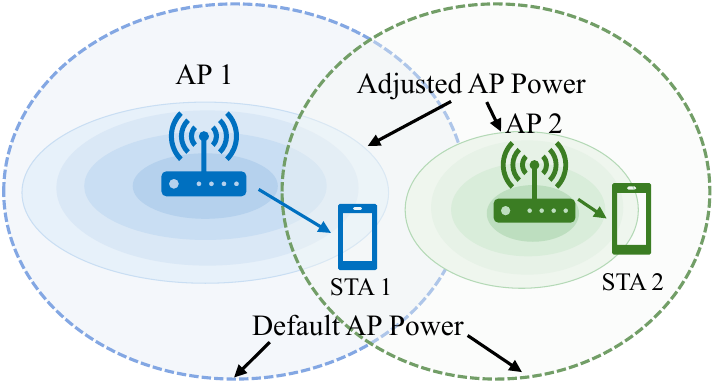}
        \label{fig:co_sr}
    }
    
    \vspace{-5pt}
    
    \subfloat[Co-BF]{
        \includegraphics[width=0.8\columnwidth]{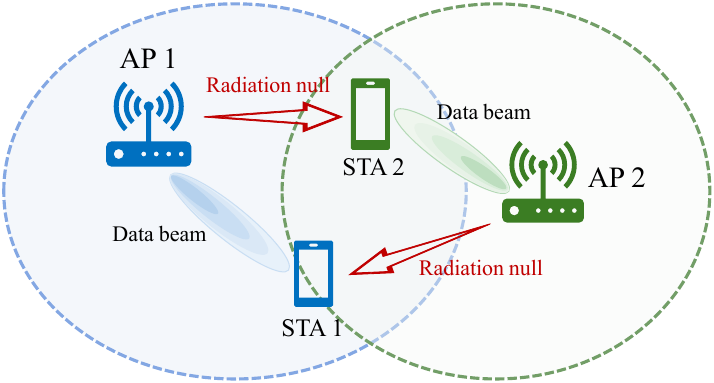}
        \label{fig:co_bf}
    }
    
    \caption{Operating principles of Co-SR and Co-BF in IEEE 802.11bn.}
    \label{fig:mapc_features}
\end{figure}

\section{Unifying Network Operations: Spatial Coordination and Seamless Mobility}
\subsection{Multi-Access Point Coordination (MAPC)}
The distributed nature of Wi-Fi networks dictates that their primary performance bottlenecks originate from inter-basic service set (BSS) interference.
This interference leads to severe channel collisions, unpredictable tail latency, and frequent packet drops at the MAC layer.
These issues become particularly pronounced in heavily loaded environments, such as stadiums, enterprise offices, and dense residential areas \cite{mapcscenarios}.
To address these fundamental challenges, the IEEE 802.11bn amendment introduces the multi-access point coordination (MAPC) framework \cite{wilhelmi2026tutorialieee80211bnmultiap, wifi25, ojcomreview}.
By transforming uncontrollable interference into precisely manageable concurrency, this mechanism allows multiple APs to collaboratively manage interference, coordinate channel access, and improve overall spectrum efficiency.

To ensure standardized coordination, the amendment defines a universal MAPC framework that governs the coordinated transmission process \cite{mapcframework}. This framework consists of four sequential phases. First, in the MAPC discovery phase, APs broadcast their coordination capabilities within management frames, such as beacons or probe response frames. This enables neighboring APs to evaluate potential coordination partners.
Second, during the MAPC negotiation phase, adjacent APs that agree to collaborate establish the specific conditions for coordinated transmissions.
This negotiation establishes essential parameters, such as their respective roles (the sharing AP and the shared AP), the maximum TXOP duration, and other related configurations.
Third, in the MAPC transmission phase, the participating APs initiate the coordinated data transmission.
This phase relies on a unified trigger mechanism where the sharing AP transmits a specific trigger frame to notify the shared AP about the exact start of the coordinated transmission. Finally, the process concludes with the MAPC teardown phase. Once the coordinated transmission is complete, any participating AP can transmit a teardown message to terminate the current coordination session.

Currently, the IEEE 802.11bn amendment primarily considers several coordination modes, including coordinated time division multiple access (Co-TDMA) \cite{ctdma}, coordinated restricted target wake time (Co-RTWT) \cite{ctwt}, coordinated spatial reuse (Co-SR) \cite{cosrframe}, and coordinated beamforming (Co-BF) \cite{cobfintro}. This article will focus specifically on the operations of Co-SR and Co-BF.
The fundamental principles of these mechanisms are illustrated in Fig.~\ref{fig:mapc_features}.

\subsubsection{Co-SR}
The core objective of Co-SR is to allow devices within OBSSs to transmit simultaneously over the same time and frequency resources under a controlled interference level \cite{cosrartical2}.
By collaboratively managing the transmit power among multiple APs, this mechanism maximizes overall network throughput and reduces queuing latency.
In traditional spatial reuse operations, a neighboring AP must reduce its transmit power to avoid interference when another AP is transmitting, which frequently degrades the signal quality for its associated STA \cite{cosrthought}.

To enable effective Co-SR, precise interference measurement and information sharing are required \cite{cosrmea1}. The sharing AP needs to understand the interference level of the other AP to cooperatively control the transmit power. A practical method involves measuring the received signal strength indicator (RSSI) between the sharing AP and the shared AP's STA, and subsequently sharing these measurements among the coordinating APs \cite{cosrmea2}.
Based on these shared measurements, the sharing AP adjusts the power level of the shared AP, ensuring an acceptable signal-to-interference-plus-noise ratio (SINR) for all participating STAs.
After determining these parameters, the sharing AP holding the TXOP transmits a trigger frame to notify the designated shared AP.
This frame initiates the coordinated transmission and indicates the duration of the shared TXOP.

Additionally, the 802.11bn standard defines two distinct modes for Co-SR.
Mode 1 allows a mixture of UHR devices and legacy EHT devices to participate, while Mode 2 is designed exclusively for pure UHR environments.
Ultimately, implementing Co-SR presents specific challenges.
Achieving this highly accurate and timely coordination among APs is particularly challenging in dense and dynamic environments where interference conditions evolve rapidly \cite{cosrartical}.

\begin{figure*}[!t]
    \centering
    \subfloat[Sequential sounding]{
        \includegraphics[width=0.95\textwidth]{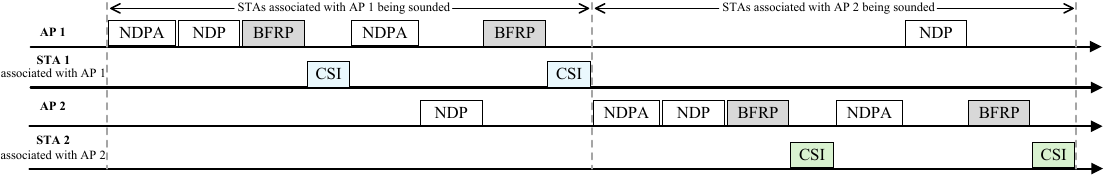}
        \label{fig:sounding_seq}
    }
    
    \vspace{-5pt}
    
    \subfloat[Joint sounding]{
        \includegraphics[width=0.88\textwidth]{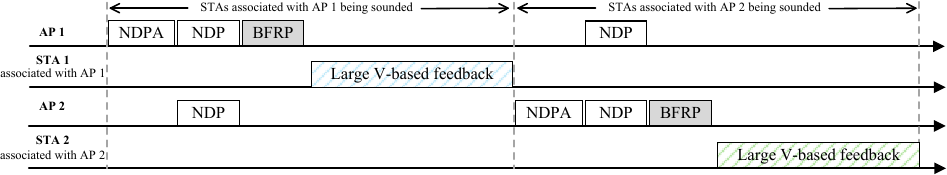}
        \label{fig:sounding_joint}
    }
    
    \caption{Comparison of sounding schemes for Co-BF. Here, BFRP denotes the beamforming report poll frame. For the joint sounding scheme, the joint NDP-based feedback will be based on large V-based feedback where the eigenvectors span the antennas across both APs.}
    \label{fig:cobf_sounding}
\end{figure*}

\subsubsection{Co-BF}
Co-BF is a spatial-domain interference cancellation technique designed to enhance the signal strength of a target STA while suppressing interference to non-target STAs \cite{cobfintro2}. Traditionally, MU-MIMO technology has been utilized to improve channel capacity, with beamforming acting as a key mechanism to mitigate intra-BSS multi-user interference. However, legacy standards only support beamforming within a single BSS. In dense deployments, uncoordinated inter-BSS interference frequently leads to severe collisions and transmission failures. To address this, the 802.11bn amendment introduces Co-BF, which leverages the spatial degrees of freedom (DoFs) inherent in MIMO systems \cite{cobfintro}. This mechanism enables an AP to transmit data to its associated STA while actively projecting spatial nulls toward STAs in adjacent BSSs \cite{cobfnull}. Consequently, Co-BF reduces collisions and enables concurrent transmissions, which significantly improves spectral reuse and lowers latency in dense deployments.

Following the MAPC framework, the Co-BF process includes request, channel sounding, and coordinated transmission phases. Among these, the channel sounding phase is the most critical, as acquiring accurate channel state information (CSI) is a prerequisite for effective beamforming \cite{cobfsounding1, cobfsounding2, cobfcsifeed}. 
As shown in Fig.~\ref{fig:cobf_sounding}, the 802.11bn standard defines two primary channel sounding methods. The first method is sequential sounding. This process is initiated by AP 1, which first transmits a null data packet announcement (NDPA) and a null data packet (NDP). This triggers STA 1 within the same BSS to perform channel measurements and feed back the CSI to AP 1. Subsequently, AP 1 sends another NDPA to instruct AP 2 to transmit an NDP frame, allowing the measurement of the CSI between AP 2 and STA 1, which is then reported back to AP 1. After AP 1 completes this entire sequence, AP 2 initiates the identical procedure to obtain the CSI associated with STA 2.
The second method is joint sounding, which allows multiple APs to simultaneously perform channel sounding for a designated STA. In this approach, AP 1 first transmits an NDPA, followed by both AP 1 and AP 2 transmitting NDPs concurrently for channel estimation. Subsequently, AP 2 repeats the same process. It is important to note that joint sounding introduces additional complexity, as it demands advanced interference cancellation algorithms at the STA side.

Once the sounding phase is completed, the coordinated transmission begins. The sharing AP transmits a Co-BF invite frame, proposing parameters such as the number of OFDM symbols, the bandwidth, and the list of participating STAs. After the shared AP replies with a response, both parties lock the final parameters via a Co-BF sync frame, thereby initiating the concurrent Co-BF transmission.

Ultimately, the performance of Co-BF depends heavily on the accuracy of the acquired CSI. If the channel conditions change frequently due to mobility or environmental dynamics, the previously calculated spatial nulls will deviate from their intended targets, bringing adverse effects to the network performance rather than improvements.

\begin{figure}[!t]
    \centering
    \includegraphics[width=\columnwidth]{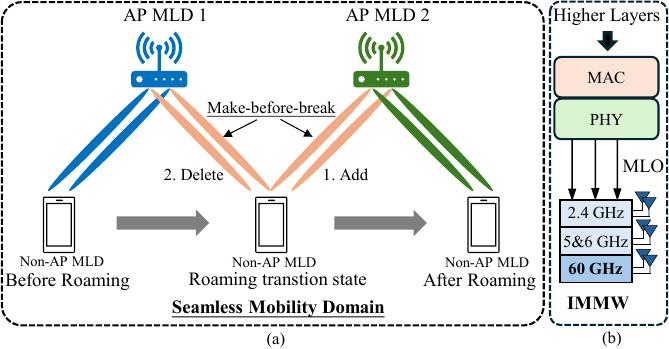}

    \vspace{-5pt} 
    
    \caption{Overview of seamless roaming and multi-link architecture. (a) Seamless roaming within a SMD. (b) Implementation of IMMW within the multi-link architecture.}
    \label{fig:seamless_roaming}
\end{figure}

\subsection{Seamless Roaming}
In traditional roaming mechanisms, a device must disconnect from the current AP before selecting and associating with a new AP, followed by re-authentication procedures.
This break-before-make approach intrinsically incurs substantial handover latency \cite{SEAMLESSdelay}.
For delay-sensitive applications such as VR and real-time industrial control, this extended roaming latency introduces severe data discontinuity.
To overcome this critical limitation, the IEEE 802.11bn task group introduces seamless roaming technology built upon the existing MLO framework \cite{SEAMLESSprocess, SEAMLESSintro}. The core objective of this technology is to allow a non-AP multi-link device (MLD) to transition smoothly between different AP MLDs within the same mobility domain.
For environments densely populated with overlapping Wi-Fi networks, this capability is essential to maintain continuous service.

The architectural foundation enabling this capability is the seamless mobility domain (SMD) \cite{SEAMLESSsmd}. The SMD integrates multiple AP MLDs within the same extended service set (ESS) into a single logical, continuous coverage area, coordinated by a unified SMD management entity.
Fig.~\ref{fig:seamless_roaming}(a) illustrates the seamless roaming process within an SMD.

The operational mechanism of seamless roaming is executed through a multi-stage process \cite{SEAMLESS1}. Initially, the non-AP MLD must discover a suitable roaming target.
This process is facilitated by embedding necessary information, such as the SMD identifier and supported transmission types, into probe response frames.
The subsequent phase is roaming preparation. Once the non-AP MLD decides to roam but remains connected to the current AP MLD, the current AP MLD and the target AP MLD preemptively transfer the operational context of the device.

Following preparation, the roaming execution phase adopts a definitive make-before-break strategy \cite{SEAMLESSprocess}. The non-AP MLD transmits a roaming request frame to the current AP MLD, specifying the identifier of the target AP MLD. 
\rev{Subsequently, the target AP MLD sends an execution response to the non-AP MLD via the current AP MLD.}
Crucially, the current AP MLD does not terminate the connection immediately. Instead, it continues to transmit buffered downlink data to the non-AP MLD for a \rev{downlink draining period indicated in the execution response, during which the target AP MLD could also transmit data to the non-AP MLD, thereby} ensuring uninterrupted data reception during the transition. 
\rev{The roaming execution is completed once the downlink draining duration expires, or either the non-AP MLD or the current AP MLD sends a notify frame to terminate the downlink draining.}
Throughout this entire procedure, the SMD architecture avoids the need for re-authentication, ensuring that the non-AP MLD achieves rapid and seamless roaming across multiple AP MLDs.

\section{Expanding Spectrum Horizons: Millimeter-Wave Integration and Cross-Band Synergy}
A primary driver of innovation in wireless communications is the continuous expansion of available spectrum. Although Wi-Fi 6E and Wi-Fi 7 extended operations into the 6 GHz band and introduced channel bandwidths up to 320 MHz, the wireless spectrum remains a finite physical resource \cite{immvintro}.
The imbalance between spectrum supply and escalating data demand is becoming increasingly acute, meaning that further expanding channel bandwidths within the traditional sub-7 GHz bands will be highly difficult.
These physical constraints restrict the capacity of future Wi-Fi systems to deliver the high throughput and low latency required by novel applications, such as AR/VR, high-precision ranging, and environmental sensing \cite{LiuIMMW}.

To overcome this spectrum scarcity, the industry is shifting its focus toward the abundant unlicensed spectrum available in the 60 GHz mmWave band \cite{WiFiAlliancewigig, immwchannelaccess}.
Utilizing the mmWave band enables high transmission rates and low communication latency.
Additionally, since a large number of independent channels can be allocated within this wide spectrum, the technology is highly suitable for operations in dense environments.
Previously, the IEEE 802.11ad \cite{80211ad} and 802.11ay \cite{80211ay} standards defined operations in the mmWave band, but their market penetration remained limited.
The main reason for this limited adoption was the requirement to develop a completely independent mmWave modem.
These earlier standards were unable to reuse \rev{the PHY hardware components of lower-frequency bands under the framework of MLO}, which inevitably resulted in complex chip designs and high research and development costs.

\begin{table*}[t]
\centering
\caption{Baseline System Parameters for UHR Performance Verification Scenarios}
\label{tab:simulation_parameters}
\renewcommand{\arraystretch}{1.2}
\setlength{\tabcolsep}{4.5pt} 
\begin{tabular}{@{} l c c c c c c c c c @{}}
\toprule
\multirow{3}{*}{\textbf{Parameters}} & \multicolumn{2}{c}{\textbf{Throughput}} & \multicolumn{5}{c}{\textbf{Latency}} & \multicolumn{2}{c}{\textbf{Packet Loss}} \\
\cmidrule(lr){2-3} \cmidrule(lr){4-8} \cmidrule(lr){9-10}
 & \multicolumn{2}{c}{\textbf{NPCA}} & \multicolumn{2}{c}{\textbf{P-EDCA}} & \multicolumn{3}{c}{\textbf{NPCA}} & \multicolumn{2}{c}{\textbf{DUO}} \\
\cmidrule(lr){2-3} \cmidrule(lr){4-5} \cmidrule(lr){6-8} \cmidrule(lr){9-10}
 & \textbf{Scen. A} & \textbf{Scen. B} & \textbf{Scen. A} & \textbf{Scen. B} & \textbf{Scen. A} & \textbf{Scen. B} & \textbf{Scen. C} & \textbf{Scen. A} & \textbf{Scen. B} \\
\midrule
\textbf{Number of BSSs} & 2 & 2 & 1 & 2 & 2 & 2 & 2 & 1 & 1 \\
\textbf{STAs per BSS} & 3 & 3 & 10 & 10 & 3 & 3 & 3 & 3 & 3 \\
\textbf{NPCA switch delay} & 50 \unit{\micro\second} & 50 \unit{\micro\second} & N/A & N/A & 50 \unit{\micro\second} & 50 \unit{\micro\second} & 50 \unit{\micro\second} & N/A & N/A \\
\textbf{Rate BSS 1 (Mbps)} & 10--1000 & 1000 & 1--100 & 1--20 & 10--1000 & 200 & 200 & 100 & 100 \\
\textbf{Rate BSS 2 (Mbps)} & 10--1000 & 50--1000 & N/A & 1--20 & 10--1000 & 40--200 & 40--200 & N/A & N/A \\
\textbf{Traffic direction} & DL & DL & UL & UL & DL & DL & UL & DL \& UL & DL \& UL \\
\textbf{AC} & AC\_BE & AC\_BE & AC\_VO & AC\_VO & AC\_BE & AC\_BE & AC\_BE & AC\_BE & AC\_BE \\
\textbf{MCS} & 11 & 11 & 9 & 9 & 11 & 11 & 11 & 11 & 11 \\
\textbf{$\boldsymbol{CW}_{\mathbf{max}}$} & 1023 & 1023 & 15 & 15 & 1023 & 1023 & 1023 & 1023 & 1023 \\
\textbf{$\boldsymbol{CW}_{\mathbf{min}}$} & 31 & 31 & 7 & 7 & 31 & 31 & 31 & 31 & 31 \\
\midrule
\textbf{Packet size} & \multicolumn{9}{c}{1500 Bytes} \\
\textbf{NSS} & \multicolumn{9}{c}{2} \\
\textbf{A-MPDUs} & \multicolumn{9}{c}{256} \\
\textbf{AIFS} & \multicolumn{9}{c}{34 \unit{\micro\second}} \\
\bottomrule
\end{tabular}
\label{tableSIM}
\end{table*}

In February 2025, the IEEE established the 802.11bq task group, formally known as the Integrated Millimeter Wave (IMMW) project \cite{IEEE80211bqPAR}.
According to its PAR, the amendment outlines non-standalone WLAN operations in the 42 GHz to 71 GHz unlicensed bands utilizing single-user (SU) OFDM transmissions.
The PAR specifies that IMMW-capable devices must support at least one sub-7 GHz band and extend the MLO framework to manage these high-frequency links.
A central directive of this project is to leverage existing PHY and MAC specifications from sub-7 GHz operations to ensure backward compatibility and minimize new protocol overhead.
Fig.~\ref{fig:seamless_roaming}(b) illustrates the architecture of IMMW enabled by MLO.

To achieve these objectives at the PHY layer, IMMW avoids designing a new control PHY to minimize chip complexity and production costs.
\rev{Taking into account the data rate requirements of most applications and the implementation costs, the IEEE 802.11bq task group currently plans to support two bandwidth modes of 320 MHz and 640 MHz \cite{immwQualcomm, immwMediatek}.}
IMMW maximizes the reuse of existing sub-7 GHz PHY architectures \cite{immw7ghz}.
\rev{Specifically, the data tone plan can be constructed by upclocking the legacy IEEE 802.11ac or 802.11be architectures \cite{immwsgdiscuss, immwsgpar}.}
Using the IEEE 802.11ac architecture as an example, a device can multiply the original channel bandwidth \rev{(including 80~MHz and 160~MHz)} and the subcarrier spacing of 312.5 kHz by a scaling factor \rev{of four}, to operate directly in the mmWave band \cite{immwQualcomm, immwMediatek, immwhuaweispace}.
During this conversion, the number and indices of the subcarriers remain unchanged, which allows the underlying baseband processing modules to be completely preserved.
\rev{IMMW plans to adopt a subcarrier spacing of 1.25 MHz, which yields higher PHY efficiency, including improved tone plan efficiency and reduced guard interval (GI) overhead \cite{immwQualcomm}.}
At the MAC layer, IMMW reuses the MLO framework introduced in 802.11be, expanding its capabilities to coordinate operations across both low-frequency and mmWave bands.
Because mmWave signals are highly susceptible to physical blockages, while sub-7 GHz signals offer strong penetration and robustness, IMMW implements a cross-band separation of the control and data planes.
Through the MLO architecture, devices route essential control and management traffic, including network discovery, association, and beacon frame transmissions, exclusively through the sub-7 GHz band \cite{immwmac}.
When data transmission is required, the AP can switch to the mmWave band via MLO specifically for high-throughput data delivery.

In summary, compared to the earlier 802.11ad and 802.11ay standards, the distinguishing characteristics of mmWave \rev{integration in the Wi-Fi 8 era} include the maximal reuse of sub-7 GHz band modems, the elimination of a standalone control PHY, and the structural integration of the MLO framework.

While mmWave technology improves throughput and reduces latency, communications in these high-frequency bands are susceptible to hardware impairments, such as carrier frequency offset (CFO), phase noise (PN), and power amplifier (PA) non-linearity \cite{immwhardware}.
Specifically, CFO introduces severe inter-carrier interference (ICI) in OFDM systems, which must be compensated by employing CFO estimation algorithms.
\rev{To support worst-case CFO correction exceeding 2.5 MHz, the STF in the IMMW PPDU is modified to populate every 8th subcarrier \cite{immwhuaweiPreamble, immwQualcomm}.}
Additionally, PN becomes more severe at high frequencies and significantly affects overall system performance, making it crucial to implement reliable phase correction mechanisms.
Furthermore, to compensate for the substantial spatial path loss inherent to mmWave transmissions, the PA must operate close to its saturation point to achieve maximum efficiency.
However, operating in this region causes non-linear signal distortion. Consequently, the system must carefully balance the amount of power backoff to mitigate distortion while maintaining adequate transmit power.

\section{Validating the UHR Mandate: System-Level Evaluations}
\subsection{Throughput Improvement Verification}

As previously outlined, the PAR for IEEE 802.11bn mandates at least one operational mode capable of achieving a 25 percent increase in MAC layer throughput.
Multiple features within the standard contribute to this capacity enhancement, including NPCA, Co-BF, Co-SR, and dynamic bandwidth expansion (DBE).
In earlier standard proposals, the industry has already demonstrated that Co-BF and Co-SR can meet this 25 percent target, and the throughput improvements derived from DBE operations are inherently straightforward.
However, there remains a lack of system-level simulations evaluating the actual throughput gains of the NPCA mechanism in complex interference environments.
Therefore, this section conducts simulations to evaluate the throughput enhancement capabilities of NPCA.

Before conducting independent analyses of different scenarios, we establish the baseline PHY and MAC parameters.
The simulated network comprises two OBSSs (BSS 1 and BSS 2), with each BSS supporting three non-AP STAs.
The generated traffic consists entirely of downlink (DL) flows with a fixed packet size of 1500 bytes.
For the transmission configurations, the MCS is set to 11, the number of spatial streams (NSS) is 2, and the maximum aggregated MAC protocol data units (A-MPDUs) limit is configured to 256.
The access category (AC) is set to best effort (AC\_BE), with a minimum contention window ($CW_\mathrm{min}$) of 31, a maximum contention window ($CW_\mathrm{max}$) of 1023, and an arbitration interframe space (AIFS) of 34 \unit{\micro\second}.
Additionally, for devices supporting the NPCA feature, the channel switching delay is specified as 50 \unit{\micro\second}.
To investigate the performance of NPCA, we designed two scenarios. Scenario A aims to verify the NPCA throughput gains when the overlapping networks possess asymmetric bandwidth capabilities. Meanwhile, Scenario B is structured to evaluate the impact of NPCA signaling overhead on overall performance when the entire channel operates under high-load conditions.
The parameter settings for different scenarios are summarized in Table~\ref{tableSIM}.

\begin{figure}[!t]
    \centering
    \subfloat[Scenario A]{
        \includegraphics[width=\columnwidth]{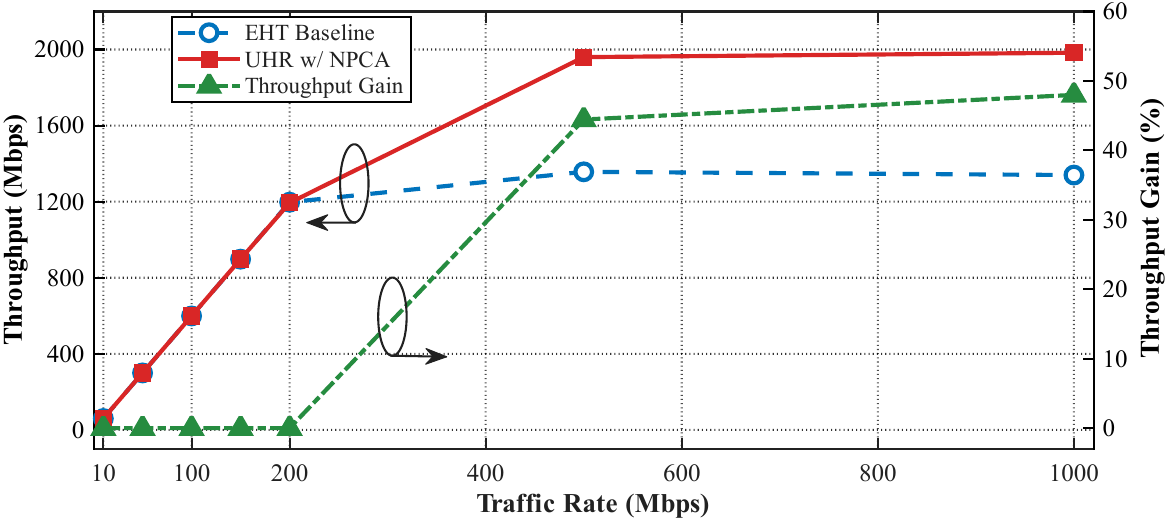}
        \label{fig:t_npca_a}
    }
    
    \vspace{-5pt}
    
    \subfloat[Scenario B]{
        \includegraphics[width=\columnwidth]{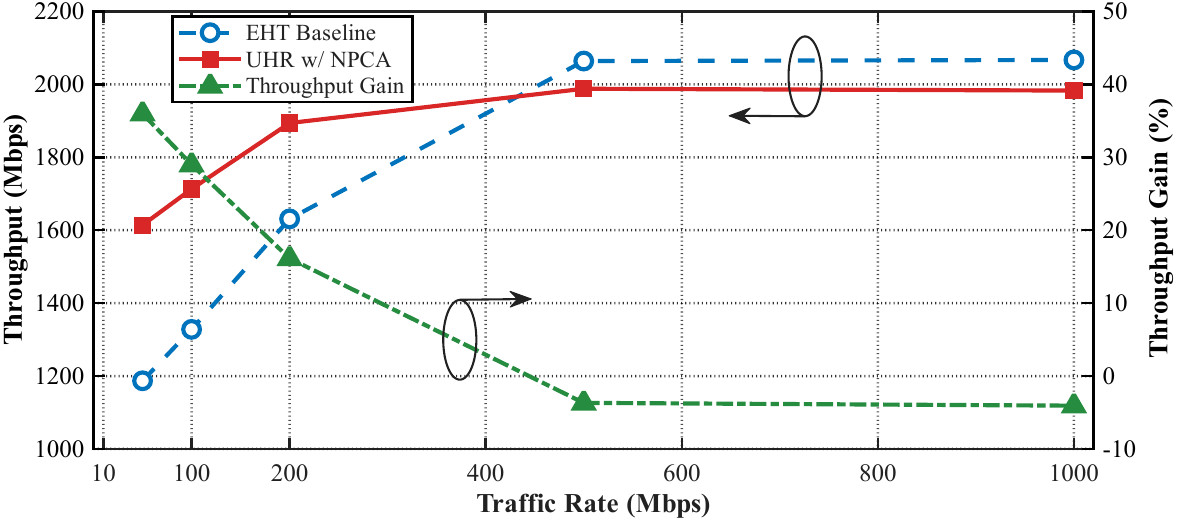}
        \label{fig:t_npca_b}
    }
    
    \caption{Simulation results of throughput enhancement with the NPCA feature.}
    \label{fig:throughput_npca}
\end{figure}

Scenario A simulates a network with mixed bandwidth capabilities. In this configuration, BSS 1 supports a 160 MHz bandwidth and enables NPCA, while BSS 2 is limited to an 80 MHz bandwidth without NPCA support. The PCHs of the two BSSs overlap. The EHT baseline maintains identical channel bandwidth settings but lacks NPCA support in both BSSs. For the traffic load, the data rate per STA in both BSSs increases synchronously from 10 Mbps to 1000 Mbps. This setup allows us to observe throughput variations as the network transitions from a light load to an overloaded state.
Fig.~\ref{fig:throughput_npca}(a) illustrates the throughput variation curves. In the low-load region, NPCA does not provide any throughput gain.
This phenomenon is logical because the total data volume remains small before network saturation.
Although BSS 1 utilizes NPCA to switch to the SCH early when the PCH is occupied, the baseline EHT mode still has sufficient time to clear its buffer queues once the PCH becomes idle. The throughput gain of NPCA becomes apparent only when the EHT simulation reaches throughput saturation. When the traffic rate reaches 500 Mbps, the PCH is densely occupied by transmissions from BSS 2, preventing BSS 1 in the EHT mode from transmitting data due to PCH blockage. However, the STAs in BSS 1 with NPCA support can immediately switch to the idle secondary 80 MHz channel upon detecting a busy PCH, allowing them to continue data transmission. The simulation results indicate that the throughput gain of NPCA can exceed 40\% in this scenario with partial spectrum overlap.

Scenario B differs from Scenario A solely in the EHT baseline configuration. \rev{While the NPCA simulation retains the overlapping setup from Scenario A (with both BSSs sharing the same PCH to generate OBSS traffic), BSS 1 and BSS 2 in the EHT baseline are set to occupy non-overlapping 80 MHz bands.}
For the traffic load, we apply an asymmetric configuration. The input traffic for BSS 1 is maintained at a saturated state of 1000 Mbps, while the traffic rate for BSS 2 serves as a variable, increasing progressively from 50 Mbps to 1000 Mbps. We designed this scenario to observe the outcome when the fully loaded BSS 1 initiates NPCA while the channel of BSS 2 becomes increasingly occupied by its own traffic.
Fig.~\ref{fig:throughput_npca}(b) presents the throughput simulation results. When the traffic rate of the STAs in BSS 2 is low, BSS 1 frequently utilizes NPCA to access that frequency band for transmission, effectively achieving data transmission over a 160 MHz bandwidth. 
In contrast, BSS 1 in the EHT simulation is restricted to an 80 MHz bandwidth.
As a result, NPCA provides a notable throughput improvement ranging from 16.125\% to 35.973\%.
However, as the traffic rate of the STAs in BSS 2 increases, the opportunities for BSS 1 to use the secondary channel decrease, causing the throughput gain of NPCA to decline.
When the traffic rate in BSS 2 exceeds 500 Mbps, NPCA produces a negative gain.
This occurs because when BSS 2 reaches throughput saturation, the entire 160 MHz bandwidth is fully utilized in both the NPCA and EHT simulations.
Despite the lack of idle spectrum, the NPCA-enabled BSS 1 continuously attempts to trigger NPCA transitions. This behavior introduces signaling overhead, such as the ICF.
Under conditions where the secondary channel is not idle, this overhead directly leads to throughput degradation.

In summary, NPCA can provide a throughput increase of more than 25 percent compared to EHT in realistic congested scenarios with partial spectrum overlap.
Simultaneously, the simulation results illustrate the potential performance degradation caused by utilizing NPCA when the non-primary channel is inherently congested.
These findings provide a reference for designing more rational NPCA mechanisms in the future.

\subsection{Latency Reduction Verification}

The PAR for IEEE 802.11bn mandates at least one operational mode capable of reducing the 95th percentile latency by 25 percent.
This section evaluates the latency reduction capabilities of the P-EDCA and NPCA mechanisms through system-level simulations.
The system parameters for the various simulation scenarios are summarized in Table~\ref{tableSIM}.

\begin{figure}[!t]
    \centering
    \subfloat[Scenario A]{
        \includegraphics[width=\columnwidth]{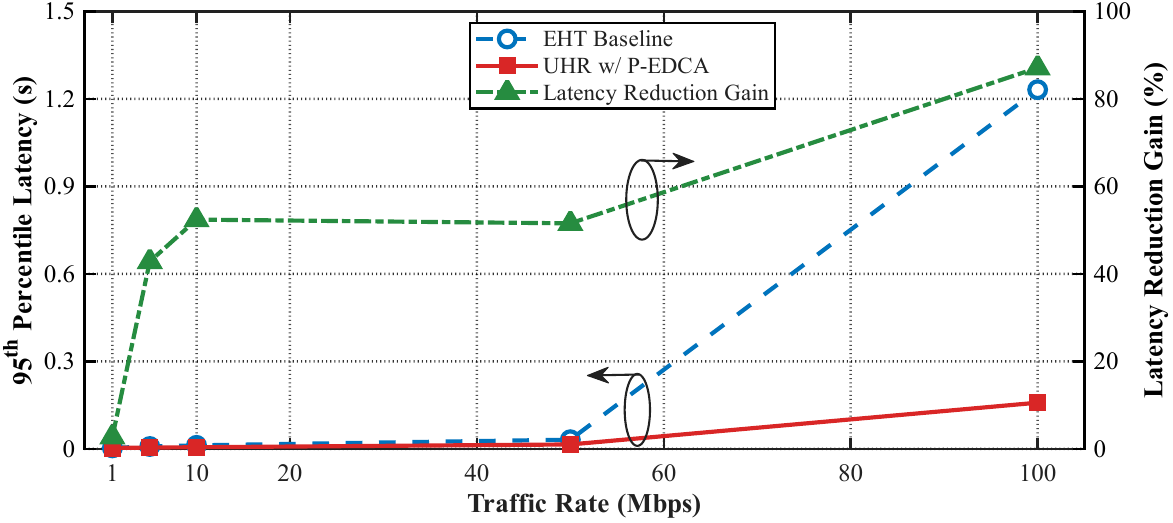}
        \label{fig:l_pedca_a}
    }
    
    \vspace{-5pt}
    
    \subfloat[Scenario B]{
        \includegraphics[width=\columnwidth]{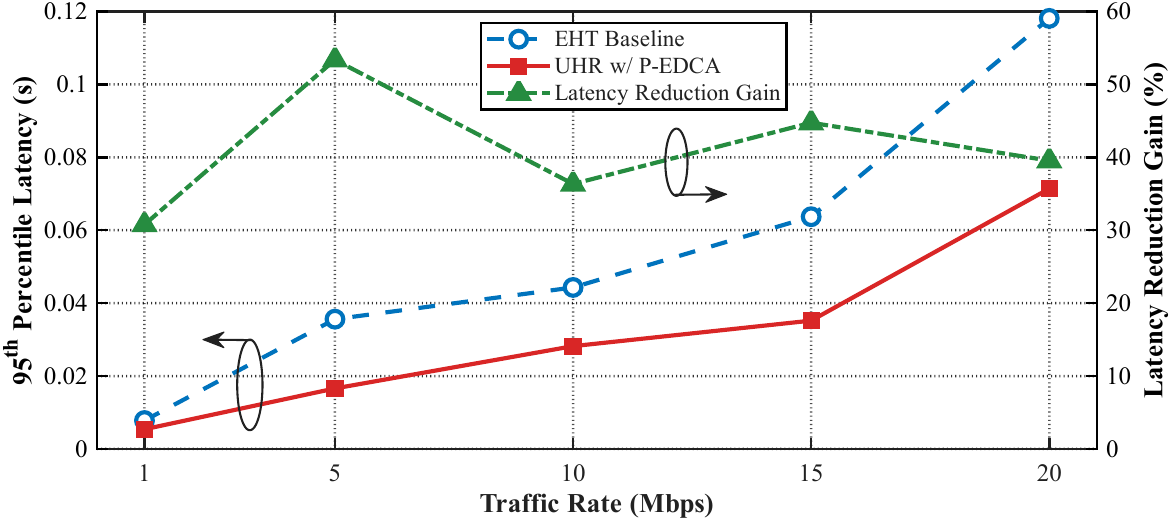}
        \label{fig:l_pedca_b}
    }
    
    \caption{Simulation results of latency reduction with the P-EDCA feature.}
    \label{fig:latency_pedca}
\end{figure}

For the P-EDCA simulations, the P-EDCA $CW_\mathrm{min}$ and $CW_\mathrm{max}$ are both configured to 7, \rev{the P-EDCA AIFS is configured with an AIFS number (AIFSN) of 2}, and \rev{the duration field of the DS-CTS frame is set to 81 \unit{\micro\second} for the protection of P-EDCA contention}. We designed two scenarios to verify the latency performance improvements enabled by P-EDCA within a single BSS and in an OBSS environment.
In Scenario A, we construct a single BSS operating over an 80 MHz bandwidth.
In this BSS, STAs 0-4 support the P-EDCA function, while STAs 5-9 lack this capability.
In the EHT baseline, none of the 10 STAs support P-EDCA.
The uplink (UL) traffic rate for each STA increases progressively from 1 Mbps to 100 Mbps.
Fig.~\ref{fig:latency_pedca}(a) presents the 95th percentile latency data for Scenario A.
The results demonstrate that at an extremely low traffic rate of 1 Mbps, the reduction in latency is minimal. However, when the traffic rate exceeds 5 Mbps, the latency reduction exceeds 40\%.
This is because when the channel is largely idle, traditional EHT devices can transmit data without enduring significant collision backoffs, making the benefits of P-EDCA less apparent.
As the traffic load increases, EHT devices experience higher latency caused by frequent channel collisions. In contrast, the P-EDCA devices significantly reduce latency through their exclusive priority mechanism. This performance advantage becomes even more evident during severe network congestion.
Based on Scenario A (BSS 1), Scenario B introduces an overlapping BSS 2, which also has an 80 MHz bandwidth and 10 associated STAs. 
In BSS 2, none of the STAs support P-EDCA. In the EHT baseline, P-EDCA remains disabled across both BSSs. As shown in Fig.~\ref{fig:latency_pedca}(b), the simulation results demonstrate that the latency reduction exceeds 30\% across all traffic rates ranging from 1 Mbps to 20 Mbps. This indicates that P-EDCA can consistently provide latency reduction even in the presence of more complex OBSS interference.

\begin{figure}[!t]
    \centering
    \subfloat[Scenario A]{
        \includegraphics[width=\columnwidth]{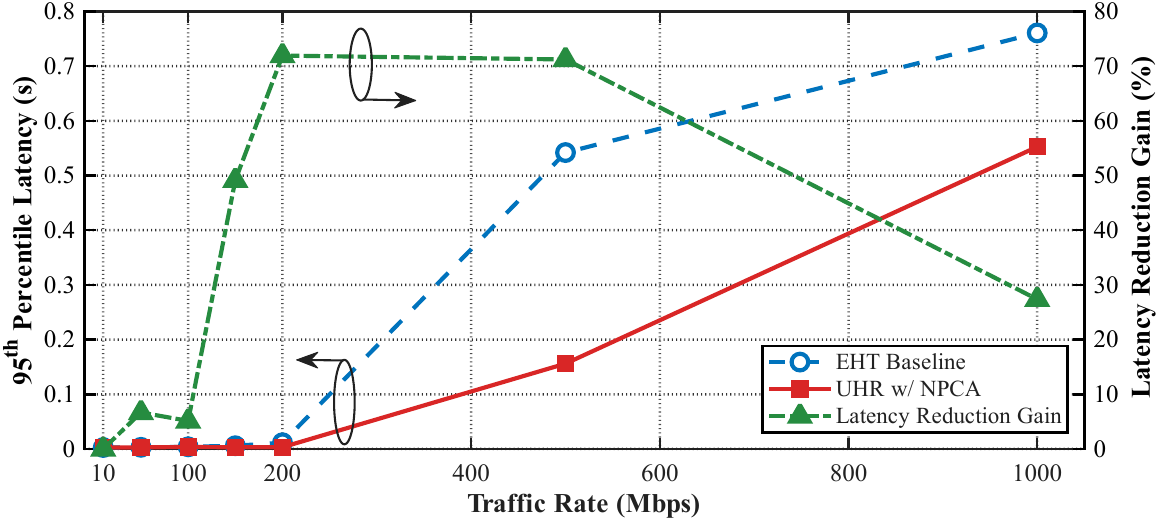}
        \label{fig:l_npca_a}
    }
    
    \vspace{-5pt}
    
    \subfloat[Scenario B]{
        \includegraphics[width=\columnwidth]{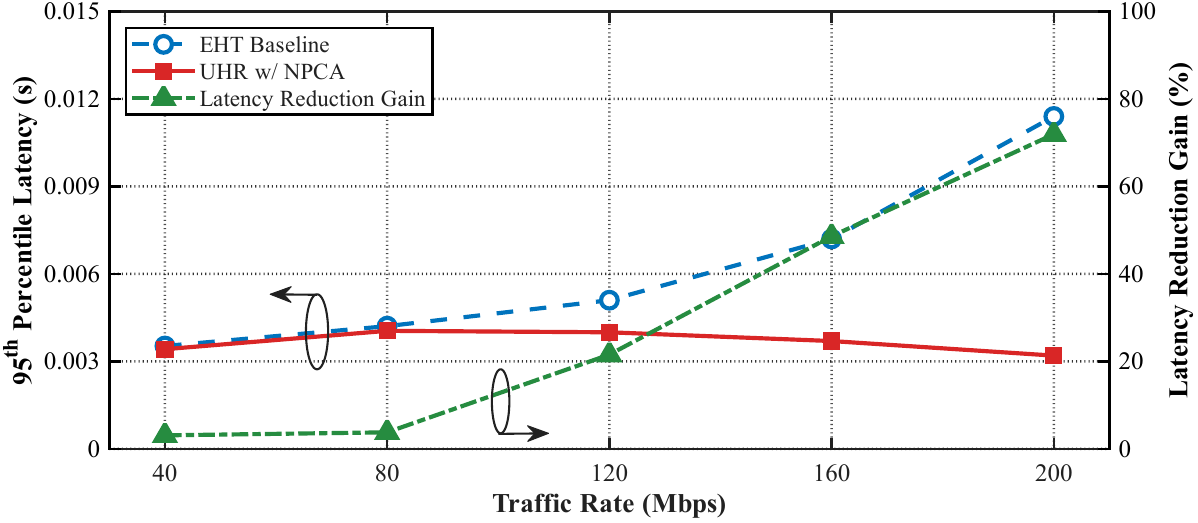}
        \label{fig:l_npca_b}
    }
    
    \vspace{-5pt}
    
    \subfloat[Scenario C]{
        \includegraphics[width=\columnwidth]{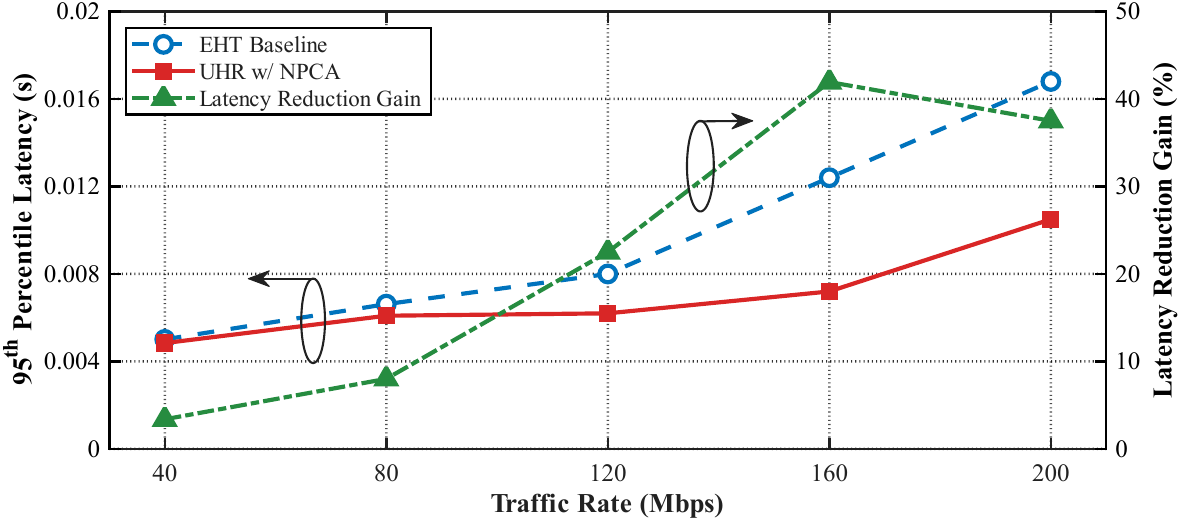}
        \label{fig:l_npca_c}
    }
    
    \caption{Simulation results of latency reduction with the NPCA feature.}
    \label{fig:latency_npca}
\end{figure}

\rev{In the simulation scenarios for NPCA}, we design a network topology comprising two OBSSs. 
BSS 1 operates with a 160 MHz bandwidth and supports NPCA, while BSS 2 operates with an 80 MHz bandwidth and does not support NPCA.
The EHT baseline utilizes the same bandwidth settings for both BSSs but lacks NPCA functionality.
In Scenario A, the traffic direction is DL, and the traffic rate per STA in both BSS 1 and BSS 2 increases synchronously from 10 Mbps to 1000 Mbps.
Fig.~\ref{fig:latency_npca}(a) illustrates the latency simulation results. At low traffic rates of 100 Mbps and below, the latency reduction achieved by NPCA is minimal.
This is because the probability of PCH occupancy remains low under light load conditions.
At moderate traffic rates (150\,Mbps--500\,Mbps), the latency reduction ranges between 49.15\% and 71.93\%.
During this phase, the PCH is occupied by BSS 2 for extended periods.
Devices supporting NPCA can transition to the idle SCH to transmit data, thereby reducing the channel access delay.
When traffic saturation occurs at 1000 Mbps, the latency reduction declines to 27.37\%. This decrease is attributed to the SCH also becoming fully occupied by the large traffic volume, which limits the overall gain of the NPCA mechanism.
Scenario B adopts an asymmetric DL load configuration.
The traffic rate for the STAs in BSS 1 is fixed at 200 Mbps, while the traffic rate for the STAs in BSS 2 increases progressively from 40 Mbps to 200 Mbps. 
As shown in Fig.~\ref{fig:latency_npca}(b), the latency reduction of NPCA compared to the EHT baseline increases as the traffic rate in BSS 2 rises.
It is noteworthy that within the NPCA simulation, the actual latency slightly decreases as the traffic rate in BSS 2 grows.
This occurs because when the traffic rate increases in BSS 2, more MPDUs are aggregated to form a longer PPDU.
Consequently, the AP in BSS 1 obtains a longer transmission window during the NPCA operation, providing it with more opportunities to transmit buffered data packets to its associated STAs.
The configuration for Scenario C is identical to Scenario B, except that the traffic direction is changed to UL.
In this scenario, channel contention becomes more complex.
The simulation results in Fig.~\ref{fig:latency_npca}(c) indicate that although the latency performance gain is lower than that of the downlink scheduled by the AP, this gain still exceeds the 25 percent target defined in the PAR.

\begin{figure}[!t]
    \centering
    \subfloat[Scenario A]{
        \includegraphics[width=\columnwidth]{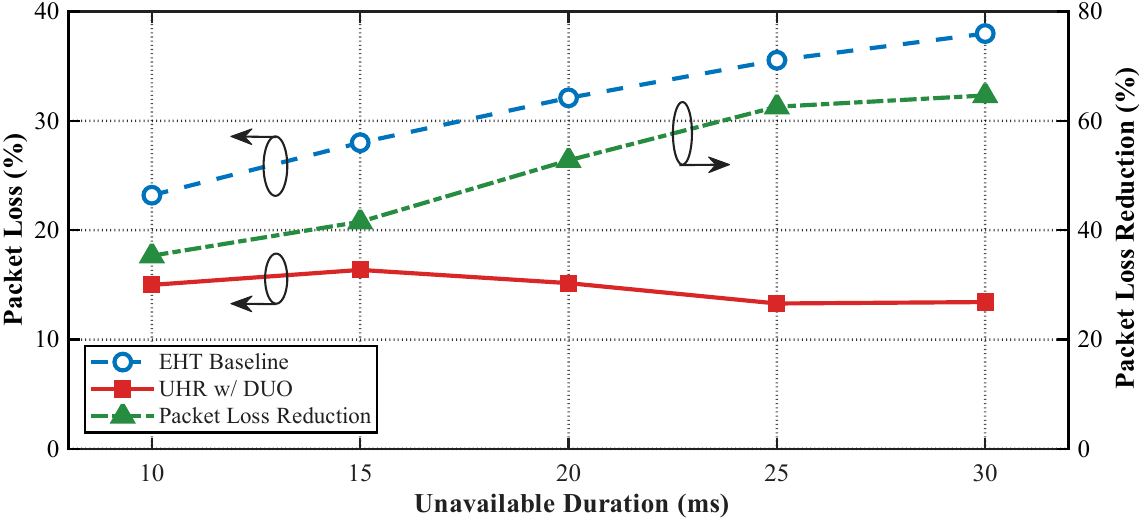}
        \label{fig:p_duo_a}
    }
    
    \vspace{-5pt}
    
    \subfloat[Scenario B]{
        \includegraphics[width=\columnwidth]{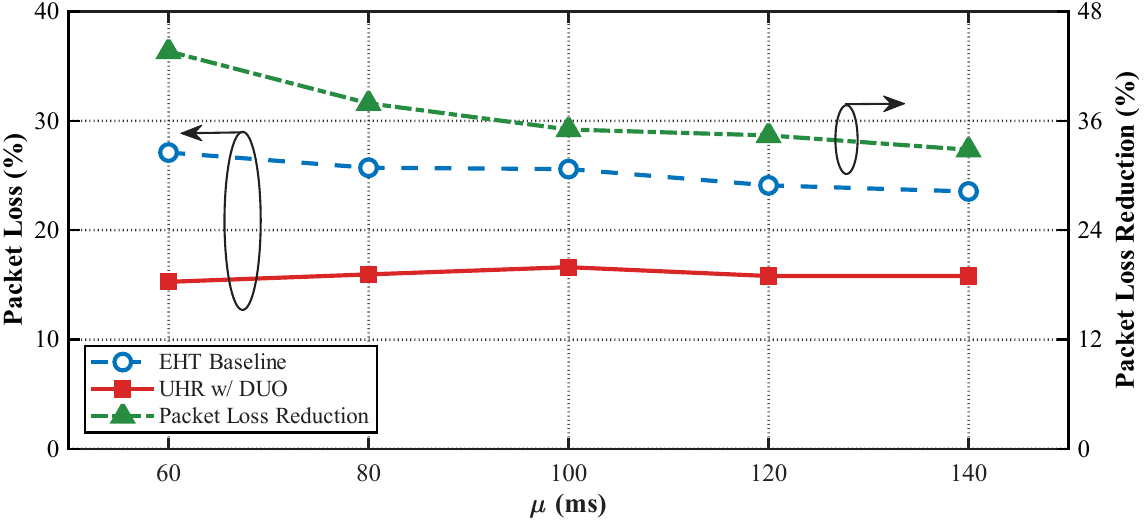}
        \label{fig:p_duo_b}
    }
    
    \caption{Simulation results of packet loss reduction with the DUO feature.}
    \label{fig:packet_duo}
\end{figure}

In summary, both P-EDCA and NPCA can effectively reduce latency. P-EDCA consistently provides a latency reduction of over 30\%, whether within a single BSS or across OBSSs. Additionally, NPCA achieves significant latency reduction under moderate and high traffic rates, whereas the latency reduction is minor under low traffic rates.

\subsection{Packet Loss Reduction Verification}

The PAR for IEEE 802.11bn mandates at least a 25 percent reduction in packet loss. 
It is difficult to directly observe packet loss during BSS transitions through simulations.
Therefore, this section evaluates the reduction of packet loss during data transmission by simulating the DUO and IM features.

For the DUO simulations, we designed two specific scenarios, with the foundational simulation parameters listed in Table~\ref{tableSIM}. In Scenario A, the average interval between the arrival times of two adjacent packets is fixed at $\mu = $ 100 ms, and DUO is applied to the non-AP STAs. The EHT baseline does not support DUO. Instead, the baseline STAs attempt to transmit a CTS frame within 1 ms before the unavailable state begins. This CTS frame can reserve the channel for a maximum duration of 4.096 ms.
As shown in Fig.~\ref{fig:packet_duo}(a), the unavailable duration is progressively increased from 10 ms to 30 ms.
The simulation results indicate that for STAs enabling DUO, the packet loss rate does not deteriorate with longer unavailable durations and even exhibits a slight decrease. This occurs because a longer unavailable period reduces the number of actively contending STAs, thereby lowering the collision probability.
Furthermore, more data packets accumulate in the buffer after the unavailable period concludes, providing more aggregated MPDUs for subsequent transmissions.
Conversely, the packet loss rate for the EHT baseline increases as the unavailable duration \rev{extends}. This degradation is primarily because the network allocation vector (NAV) protection provided by the CTS frame cannot cover the entire unavailable period. In addition, some STAs might fail to successfully transmit the CTS frame before the unavailable period begins. Compared to the EHT baseline, the application of DUO consistently achieves a packet loss reduction exceeding 35\%.
In Scenario B, the unavailable duration of the devices is fixed at 10 ms, and $\mu$ serves as the variable. 
In the simulation, $\mu$ increases from 60 ms to 140 ms, representing a transition from dense to sparse network traffic.
The EHT baseline settings remain identical to those in Scenario A. As illustrated in Fig.~\ref{fig:packet_duo}(b), the packet loss rate under the DUO mechanism shows no significant variation across different traffic loads and consistently provides a packet loss reduction of over 30\%.
This demonstrates that DUO effectively eliminates packet loss during the unavailable periods of the STA.
For the EHT baseline, the packet loss rate exhibits a slight decrease as $\mu$ increases.
This reduction stems from the lower probability of a data transmission coinciding with an unavailable period under sparse traffic conditions.

\begin{figure}[!t]
    \centering
    \includegraphics[width=\columnwidth]{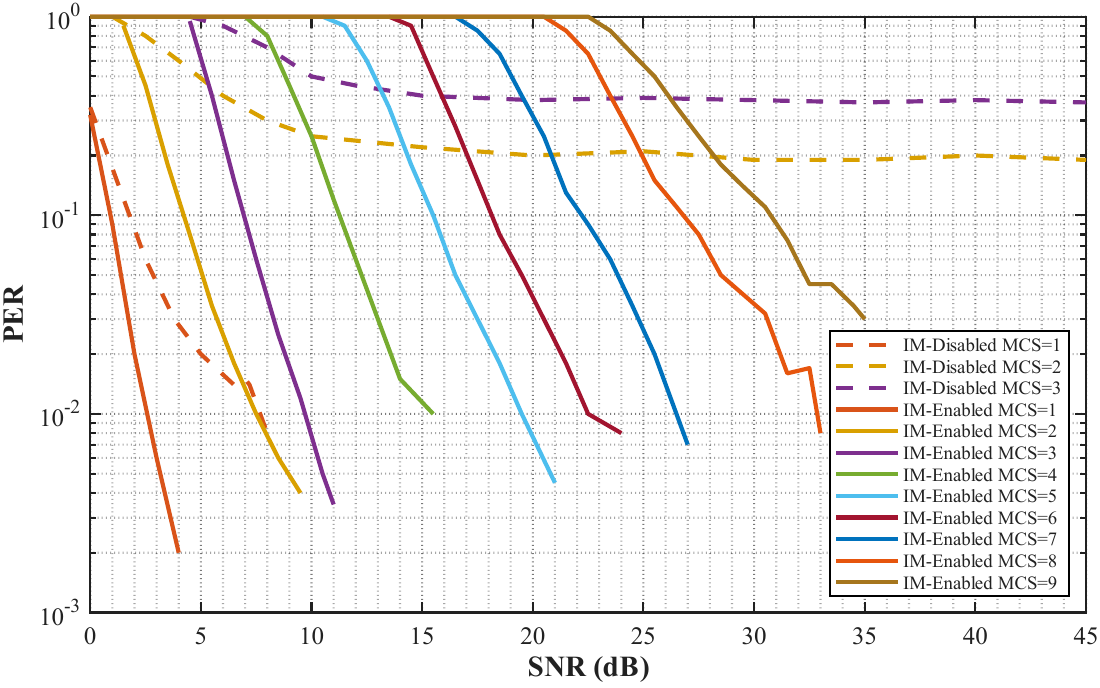}
    \caption{Simulation results of the impact of the IM feature on PER.}
    \label{fig:im_impact}
\end{figure}

In the IM feature simulation, we construct a scenario characterized by severe interference. The transmitter is equipped with two antennas, and the receiver is equipped with four antennas, while the system transmits only a single spatial data stream. Additionally, we introduce one interference stream and set the signal-to-interference ratio (SIR) to 5 dB. Fig.~\ref{fig:im_impact} illustrates the packet error rate (PER) simulation results across different MCSs with the IM mechanism enabled and disabled. The results show that when IM is disabled, an error floor emerges at MCS 2 and above because the external interference is too strong. After enabling the IM mechanism, the receiver possesses redundant spatial degrees of freedom since the number of receive antennas exceeds the number of spatial streams. Consequently, the PER decreases as the signal-to-noise ratio (SNR) increases, and the system successfully supports modulation formats up to MCS 9. This demonstrates that when the transmitter selects a higher MCS, the application of IM achieves a significant packet loss improvement compared to the EHT baseline.

In summary, the application of DUO achieves an MPDU loss reduction of more than 30\% across various scenarios. Furthermore, in strong interference environments, IM provides a significant MCS improvement, thereby effectively reducing the packet loss rate.

\begin{figure*}[!t]
    \centering
    \subfloat[PN only.]{
        \includegraphics[width=0.31\textwidth]{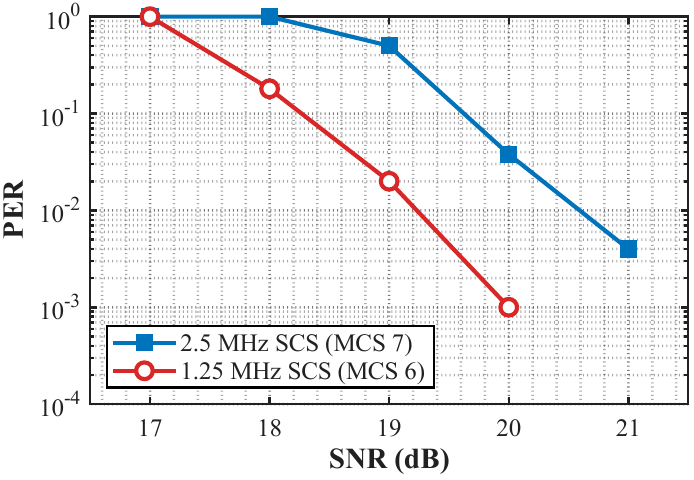}
        \label{fig:scs_a}
    }\hfill%
    \subfloat[PA non-linearity only.]{%
        \includegraphics[width=0.31\textwidth]{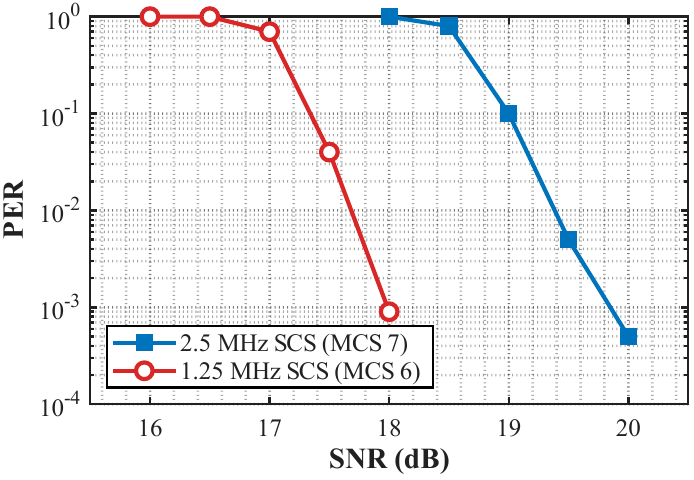}
        \label{fig:scs_b}
    }\hfill%
    \subfloat[Both PN and PA non-linearity.]{%
        \includegraphics[width=0.31\textwidth]{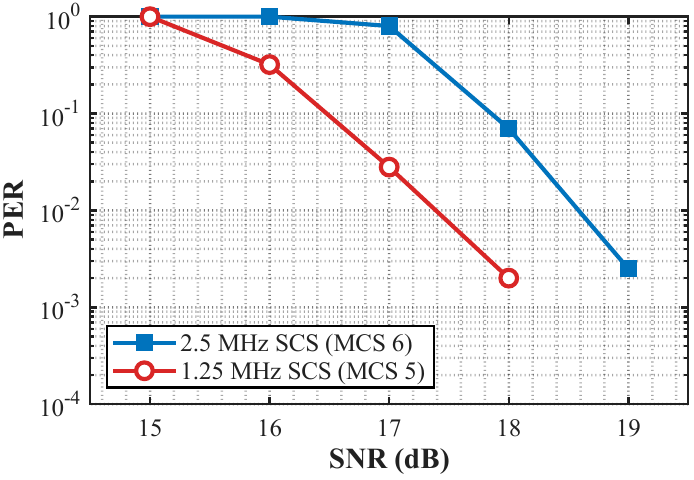}
        \label{fig:scs_c}
    }
    
    \caption{PER performance comparison between the 1.25 MHz and 2.5 MHz subcarrier spacing (SCS) schemes under hardware impairments.}
    \label{fig:scs_comparison}
\end{figure*}

\subsection{Integrated Millimeter Wave Performance Verification}
We conducted system-level simulations to verify the feasibility of reusing the sub-7 GHz PHY architecture and translating the signal directly to the mmWave band through upclocking.
\rev{Specifically, we verified through simulations the system-level performance advantages of adopting a 1.25 MHz subcarrier spacing (SCS) in the presence of hardware impairments.}
\rev{The simulation utilizes an upclocked PPDU based on IEEE 802.11ac and incorporates the effects of PN and PA non-linearity.}

\rev{Regarding the communication architecture, the simulation adopts a single-input single-output (SISO) antenna configuration operating over a 320 MHz bandwidth. For comparison, we establish two implementation schemes: upclocking the IEEE 802.11ac 80 MHz architecture by a factor of four, and upclocking the 40 MHz architecture by a factor of eight. These approaches yield subcarrier spacings of 1.25 MHz and 2.5 MHz, respectively.}
\rev{The 80 MHz and 40 MHz PPDUs reuse the very high throughput (VHT) format of IEEE 802.11ac, with a fixed packet length of 2048 bytes.}
\rev{Furthermore, the system employs LDPC coding and a GI of 50 ns, with the MCS set to 5, 6, and 7.}
Additionally, the simulation uses the conference room (CR) channel model specified in IEEE 802.11ad.
For the hardware impairments, we establish specific configurations. 
\rev{For the PN, we utilize a one-pole and one-zero phase noise model \cite{immvphasemodel}, where the pole frequency is set to 1 MHz, the zero frequency is configured at 100 MHz, and the PSD zero point is specified as $-$90~dBc/Hz.
To model the PA non-linearity, we apply a modified Rapp model \cite{immwrapp}.}

\rev{Fig.~\ref{fig:scs_comparison} illustrates the PER performance comparison between the 1.25 MHz and 2.5 MHz subcarrier spacing schemes under various hardware impairments.
Fig.~\ref{fig:scs_comparison}(a) presents the PER performance when solely considering the impact of PN.
Both schemes achieve a data rate of approximately 1.2 Gbps.
Benefiting from higher PHY efficiency, the 1.25 MHz scheme only needs MCS 6, whereas the 2.5 MHz scheme requires MCS 7.
The simulation results indicate that although the 1.25 MHz subcarrier spacing theoretically experiences more severe signal degradation induced by PN, the utilization of a lower-order MCS enables it to deliver a SNR gain of approximately 1.5 dB at the same data rate.
Fig.~\ref{fig:scs_comparison}(b) evaluates the performance when solely considering PA non-linearity, with the PA backoff set to 13.5 dB.
At the same target data rate of 1.2 Gbps, although the 1.25 MHz scheme exhibits a higher PAPR due to a larger number of subcarriers, the adoption of a lower-order MCS still yields an SNR gain of 2.0 dB over the 2.5 MHz scheme.
Finally, Fig.~\ref{fig:scs_comparison}(c) presents the performance when simultaneously considering both PN and PA non-linearity, with the PA backoff set to 19 dB and the target data rate to approximately 1.1 Gbps. The 1.25 MHz scheme employs MCS 5, whereas the 2.5 MHz scheme requires MCS 6. The simulation results indicate that the 1.25 MHz scheme still delivers an SNR gain of 1.3 dB.}

In summary, the simulations confirm the feasibility of the IEEE 802.11bq IMMW architecture. 
\rev{At the same data rate, the 1.25 MHz subcarrier spacing scheme consistently delivers an SNR gain under various hardware impairments, demonstrating that it is the preferred parameter configuration for IMMW.}
\rev{Ultimately, mmWave serves as a highly promising complementary technology in the Wi-Fi 8 era.}

\section{Conclusion}
Wi-Fi 8 marks a critical turning point in the evolution of Wi-Fi technologies. Unlike previous generations that primarily pursued peak network throughput, the core vision of Wi-Fi 8 is to achieve ultra-high reliability (UHR).
This shift signifies that future wireless local area network (WLAN) designs must ensure more reliable and deterministic communication capabilities in complex and non-ideal environments.
Consequently, this transition provides a strong impetus for emerging applications such as the Industrial Internet of Things (IIoT), immersive communications, and collaborative robotics.

This paper has reviewed the evolution of Wi-Fi standards and provided a comprehensive survey of the key technological advancements in Wi-Fi 8.
Specifically, we focused on the core mechanisms of the IEEE 802.11bn PHY and MAC layers, encompassing link robustness enhancements, channel access improvements, and multi-access point coordination.
Additionally, we discussed integrated millimeter-wave (mmWave) technology and analyzed its potential role in extending the spectrum capability \rev{in the Wi-Fi 8 era} through baseband multiplexing and multi-link operation (MLO).
Unlike existing surveys that predominantly focus on protocol descriptions, a major contribution of this work is the detailed performance validation of these technologies through system-level simulations.
We verified the effectiveness of key IEEE 802.11bn features in improving throughput, latency, and packet loss performance. Furthermore, we demonstrated the robust performance of the integrated mmWave scheme even when accounting for hardware impairments.

As the standardization process for Wi-Fi 8 continues to advance, the proposed key technologies still face various practical challenges. We expect that the technical survey and system-level validation results presented in this paper will serve as a valuable reference for the ongoing research, standardization, and future real-world deployment of Wi-Fi 8 networks.

\bibliographystyle{IEEEtran}

\bibliography{IEEEabrv,references}

\vfill

\end{document}